\documentclass[12pt]{article}  % version No 4 commençée le 30 oct. 05
\def\rv{{\bf r}}
\def\vv{{\bf v}}

\def\bv{{\bf b}}

\def\kv{{\bf k}}
\def\jv{{\bf j}}
\def\sv{{\bf s}}
\def\Sv{{\bf S}}

\def\Jv{{\bf J}}
\def\Lv{{\bf L}}

\def\Pv{{\bf P}}
\def\Qv{{\bf Q}}
\def\nv{\hat{\bf n}}
\def\plan{\hat{\bf \pi}}

\def\xu{\hat{\bf x}}

\def\zu{\hat{\bf z}}

\def\ni{\noindent}

\def\tw{\tilde w}
\def\tv{\tilde v}
\def\lambdabar{\lambda\raise0.4ex\hbox{\kern-0.5em\hbox{--}}\ }
\def\lambdaC{\lambda\raise0.5ex\hbox{\kern-0.5em\hbox{--}}_{\rm C}}
\def\lesssim{{\lower0.5ex\hbox{$\stackrel{<}{\sim}$}}}
\def\gtrsim{{\lower0.5ex\hbox{$\stackrel{>}{\sim}$}}}
\begin{document}

%NOV 2005

\date{November 14, 2005}

\title
{The relativistic hydrogen atom : \\
a theoretical laboratory for structure functions%
\footnote{Presented at International Workshop on Transverse Polarisation 
Phenomena in Hard Processes - Como (Italy), 7-10 September 2005}
}

%\author{X .Artru$^{(a)}$, K. Benhizia$^{(b)}$}
\author{X. Artru%
\footnote
{Institut de Physique Nucl\'eaire de Lyon, 
IN2P3-CNRS and Universit\'e Claude Bernard, 
F-69622 Villeurbanne, France. 
E-mail: x.artru@ipnl.in2p3.fr 
}, K. Benhizia%
\footnote{Laboratoire de Physique Math\'ematique et Physique Subatomique, 
D\'epart. de Physique, Facult\'e de Sciences, 
Universit\'e Mentouri, Constantine, Algeria. 
E-mail: Beni.karima@laposte.net
} 
}

\maketitle

\vskip 1.true cm

%\centerline{X .Artru$^{(a)}$, K. Benhizia$^{(b)}$}

%\vskip 1.true cm

\vskip 1.true cm

\centerline{ABSTRACT}

\bigskip

\ni
Thanks to the Dirac equation, the hydrogen-like atom at high $Z$  
offers a precise model of relativistic bound state, allowing to 
test properties of unpolarized and polarized structure functions 
analogous to the hadronic ones, in particular: 
Sivers effect, sum rules for the vector, axial, tensor charges
and for the magnetic moment, positivity constraints, sea 
contributions and fracture functions.%

\newpage

\section{INTRODUCTION}

In this work we will study the hydrogen-like atom, 
treated by the Dirac equation, as a precise model of relativistic 
two-particle bound states when one of the constituent
is very massive. We consider the case of large $Z$ so that 
$Z \alpha \sim 1 $ and relativistic effects are enhanced. 
What we neglect is 
\begin{itemize}
\item the nuclear recoil
\item the nuclear spin (but not necessarily the nuclear size)
\item radiatives corrections, 
e.g. the Lamb shift $\sim \alpha (Z\alpha)^4 m_e$.

\end{itemize}
In analogy with the quark distributions, we will study: 
\begin{itemize}
\item the electron densities $ q (k^+) $, $ q (\kv_T,k^+) $ or
$ q (\bv, k^+ ) $ where  $ \bv = (x,y)$ is the impact parameter;
\item the corresponding polarized densities; 
\item the sum rules for the vector, axial and tensor charges
and for the atom magnetic moment;
\item the positivity constraints;
\item the electron - positron sea and the fracture functions.
\end{itemize}
We use $ k^+ = k^0 + k_z$ instead of the Bj\"orken scaling variable
$ k^+ / P^+ _{atom} $ which would be of order $ m_e/ M_{atom} $, 
hence very small.
$ | k^+ | $ can run up to $ M_{atom} $,
but typically $ | k^+ - m| \sim Z\alpha m $.

\section{Deep-inelastic probes of the electron state}

Deep-inelastic reactions on the atom are, for instance:
\begin{itemize}
\item
Compton scattering: $ \gamma_i + e^- $(bound) $ \to \gamma_f + e^-_f $(free),
\item
Moeller or Bhabha scattering (replacing the $\gamma$ by a $e^{\pm} $),  
\item
annihilation:
$ e^{+} + e^- $(bound) $ \to \gamma + \gamma$, 
\end{itemize}
the Mandelstam invariants $\hat s $, $\hat t $ and $ \hat u $
being large compared to $ m_e^2 $.
In Compton scattering for instance, 
taking the $z$ axis along $\Qv = \kv_f^{\gamma} -  \kv_i^{\gamma} $, 
the particles $ \gamma_i $, $ \gamma_f $
and $ e^-_f $ move almost in the $-z$ direction and we have 
\begin{equation}
k^+ \simeq Q^+
\,,
\label{k+simeq}
\end{equation}
\begin{equation}
\kv_T \ \simeq \ \kv_{fT} ^{e} \ + \ \kv_{fT} ^{\gamma} 
\ - \ \kv_{iT} ^{\gamma} 
\ = \ - \Pv_{fT}(nucleus) 
\,.
\label{kT} 
\end{equation}

\section{Joint $(\bv,k^+)$ distribution}
In the "infinite momentum frame" $P_z \gg M$, deep inelastic scattering
measures the gauge-invariant {\it mechanical} longitudinal momentum 
\begin{equation}
k_z \ = \ (P_z /M_{atom}) \ k^+ _{rest \ frame} \ = \ p_z - A_z 
\label{meca_z}
\end{equation}
and not the canonical one $p_z = -i \partial/\partial_z $. 
It is not possible to define a joint distribution $q(\kv_T,k_z)$
in a gauge invariant way, since the three quantum operators 
$k_i = -i\partial_i - A_i(t,x,y,z)$ are not all commuting. 
On the other hand we can define unambiguously the joint distribution 
$q(\bv,k_z)$. Note that $q(\bv,k^+)$ can be measured in atom + atom 
collisions where two hard sub-collisions occur simultaneously: 
$\{ e_1^- + e_2^- \ and \ N_1 + N_2 \}$ or: 
$\{ e_1^- + N_2 \ and \ e_2^- +  N_1 \}$. 

Given the Dirac wave function $\Psi(\rv)$ at $t=0$ 
in the atom frame, we have 
\begin{equation}
dN_{e^-}  /  [ d^2\bv \ dk^+ / (2\pi) ] \ \equiv
\ q(\bv,k^+) \ = \ \Phi^\dagger(\bv,k^+) 
\ \Phi(\bv,k^+)
\,,
\label{dens(b)} 
\end{equation}
\begin{equation}
\Phi (\bv,k^+) =  \int dz \ 
\exp\left\{-i k^+ z +i E_n z -i \chi(\bv,z) 
\right\} 
\ \Phi ( \rv ) 
\,,
\label{Phibk+}
\end{equation}
\begin{equation}
\Phi (\rv) =  
%\begin{pmatrix} %????
\pmatrix{
\Psi_1 (\rv) + \Psi_3 (\rv)  \cr 
 \Psi_2 (\rv) - \Psi_4 (\rv)  \cr 
}
%\end{pmatrix}
\,,
\label{Phi}
\end{equation}
\begin{equation}
\chi(\bv,z) = \int_{z_0}^z dz' \ V(x,y,z') 
= -Z\alpha \ \left[ \sinh^{-1} 
\left( {z \over b} \right) - \sinh^{-1} \left( {z_0 \over b} \right) \right]
\,.
\label{chi}
\end{equation}
The two-component spinor $\Phi$ represents $(1+\alpha_z)\Psi$. 
The "gauge link" $ \exp \{ -i \chi ( \bv,z ) \} $ 
transforms $\Psi$ in the Coulomb gauge to $\Psi$ in the gauge 
$A^+ = 0$ ($A_z = 0$ in the infinite momentum frame).
The choice of $z_0$ corresponds to a residual gauge freedom. 

One can check that (\ref{dens(b)}) is invariant under the 
gauge transformation $V(\rv) \to V(\rv) +$ Constant, 
$E_n \to E_n + $ Constant.
Such a transformation may result from the addition of electrons 
in outer shells; it does not change the mechanical 4-momentum 
of a K-shell electron. 

\section{Joint $(\kv_T,k^+)$ distribution}

\ni 
The amplitude of the Compton reaction is given, modulo $\alpha$ matrices, by
\begin{equation}
\langle \Psi_f | e^{-i \Qv \cdot \rv } | \Psi_i \rangle
\propto
\int d^3 \rv \ %\exp\left\
e^{ -i \Qv \cdot \rv -i \kv_f \cdot \rv -i \chi(\bv,z) } %\right\} 
\ \Phi ( \rv ) 
= \Phi (\kv_T,k^+) 
\label{ampli}
\end{equation}
with $z_0 = - \infty$.  
$\ \exp \{ -i \kv_f \cdot \rv -i \chi \} $ 
is the final wave function distorted by the Coulomb potential,
in the eikonal approximation. Equivalently,
\begin{equation}
\Phi ( \kv_T , k^+ ) 
= \int d^3 \rv \ %\exp\left\{ 
e^{ -i \kv_T \cdot \bv - i k^+ z -i \chi(\rv) } %\right\} 
\ \Phi ( \rv ) 
= \int d^2 \bv \ e^{ - i \kv_T \cdot \bv } \ \Phi(\bv,k^+)
\,,
\label{b->kT}
\end{equation}
with the identification $
k^+ = E_n + k_{f,z} + Q_z $.
[{\it check:} in a semi-classical approach,
$k^0 (\rv) = E_n - V(\rv) $ and $k_{f,z} + Q_z 
= k_{z} (\rv) - V(\rv) $ at the collision point]. 
For the annihilation reaction, %analogue to the Drell-Yan one,
the {\it incoming positron} wave is distorted, then $z_0 = + \infty$.
Thus, the gauge link takes into account either an initial or a
final state interaction~\cite{Brodsky,Yuan,Collins}. The quantity
\begin{equation}
q ( \kv_T , k^+ ) =  
\Phi^\dagger ( \kv_T , k^+ )  \ \Phi ( \kv_T , k^+ )  
\end{equation}
depends on the deep inelastic probe, contrary to $q ( \bv , k^+ ) $
and 
\begin{equation}
q ( k^+ ) = \int q ( \bv , k^+ ) \ d^2\bv 
= \int q ( \kv_T , k^+ ) \ {d^2\kv_T / (2 \pi)^2} 
\,. 
\end{equation}

\section{Spin dependence of the electron density}

Ignoring nuclear spin, the atom spin is 
$\jv = \Lv + \sv$. We will consider a $j = 1/2$ state
and denote by $ \bar \Sv^A = 2 \langle \jv \rangle $ 
and $ \bar \Sv^e = 2 \langle \sv \rangle $ 
the atom and electron polarization vectors. 
%$| \bar \Sv^A | \le 1$, $| \bar \Sv^e | \le 1$. 
$\Sv^A $ and $\Sv^e $ without bar indicate pure spin states.
The {\it unpolarized} electron density in $(\bv,k^+ )$ space 
from a {\it polarized} atom is 
\begin{equation}
q (\bv,k^+ ; \Sv^A ) \ = \ 
\Phi^\dagger(\bv,k^+ ; \Sv^A ) \ \ \Phi(\bv,k^+ ; \Sv^A )
\label{dens}
\end{equation}
and the electron polarisation is given by
\begin{equation}
\bar \Sv^e (\bv,k^+ ; \Sv^A ) \ \ q (\bv,k^+ ; \Sv^A ) \ = \  
\Phi^\dagger(\bv,k^+ ; \Sv^A ) \ \ \vec\sigma \ \ \Phi(\bv,k^+ ; \Sv^A )
\,.
\label{spin-dens}
\end{equation}

Taking into account the conservations of parity and angular momentum,
the fully polarized density 
%of electrons of spin orientation $\Sv^e$ 
%in an atom of polarisation $ \bar \Sv^A  $ 
can be written as
$$ %\begin{equation}
q (\bv,k^+, \Sv^e  ; \Sv^A) = q (b,k^+) \ [ 1 + 
C_{0n} \ (\Sv^A  \cdot \nv) +
C_{n0} \ (\Sv^e \cdot \nv) +
C_{nn} \ (\Sv^e \cdot \nv)  ( \Sv^A  \cdot \nv) 
$$ %\end{equation}
\begin{equation}
+ \ C_{ll} \  S^e_z  S^A_z  +
C_{l \pi } \ S^e_z  (\Sv^A  \cdot \plan) \ +
C_{ \pi l} \ (\Sv^e \cdot \plan) \ S^A_z  +
C_{\pi\pi} \ (\Sv^e \cdot \plan) (\Sv^A  \cdot \plan) ]  
\label{densPol}
\end{equation}%
where $\plan = \bv / b $ and $\nv = \zu \times \plan $. 
%form a basis of transverse vectors.
The $ C_{i,j} $'s are functions of $ b $ and $k^+$. 

Similar equations work for $\kv_T$ instead of $\bv$.
The link with Amsterdam notations~\cite{Bacchetta-BHM} is,
omitting kinematical factors,
\begin{eqnarray} 
q(k_T,k^+ ) &=& f_1 
\qquad \phantom{ - h_{1T}^\perp }
f_1  \ C_{0n} = f_{1T}^\perp \\
f_1  \ C_{ll} &=& g_1 
\qquad \phantom{ - h_{1T}^\perp }
f_1  \ C_{n0} = - h_1^\perp \\
f_1  \ C_{nn} &=& h_1 - h_{1T}^\perp 
\qquad
f_1  \ C_{l\pi} = g_{1T} \\
f_1  \ C_{\pi\pi} &=& h_1+h_{1T}^\perp 
\qquad
f_1  \ C_{\pi l} = h_{1L}^\perp
\,, 
\end{eqnarray} 

The $\bv$ - or $\kv_T$ integration washes out all correlations 
but $C_{ll}$ and $ C_{TT} = 
{1\over2} \left[ C_{nn} + C_{\pi\pi} \right] $, giving  
\begin{equation}
q (k^+, \Sv^e  ; \bar \Sv^A)  
= q (k^+) \ + \ \Delta q (k^+) \  S^e_z \bar S^A_z  
\ + \ \delta q (k^+) \ \ \Sv^e_T \cdot \bar \Sv^A_T 
\,.
\label{k+distrib}
\end{equation}

\subsection{Sum rules }
Integrating (\ref{k+distrib}) over $k^+$, 
one obtains the vector, axial and tensor charges
\begin{equation}
q = \int_{-\infty}^{\infty} q(k^+) \ { dk^+ / (2 \pi) } 
= \int d^3 \rv \ 
\Psi^\dagger(\rv ; \Sv^A ) \ \Psi(\rv ; any \ \Sv^A ) 
\label{V-charge}
\end{equation}
\begin{equation}
\Delta q = \int \Delta q(k^+) \ { dk^+ / (2 \pi) } 
= \int d^3 \rv \ 
\Psi^\dagger(\rv ; \Sv^A ) \ \Sigma_z \ \Psi(\rv ; \Sv^A = \zu)
\label{A-charge}
\end{equation}
\begin{equation}
\delta q = \int \delta q(k^+) \ { dk^+ / (2 \pi) } 
= \int d^3 \rv \ 
\Psi^\dagger(\rv ; \Sv^A ) \ \beta \ \Sigma_x \ \Psi(\rv ; \Sv^A = \xu)
\,.
\label{T-charge}
\end{equation}
For the hydrogen ground state, 
\begin{equation}
q = 1
\,,\qquad 
\Delta q = (1- \xi^2 /3) / (1 + \xi^2)
\,,\qquad 
\delta q = (1+ \xi^2 /3) / (1 + \xi^2)
\label{crise}
\end{equation}
where 
$\xi = Z\alpha /(1+\gamma)$, $\ \gamma = E/m_e = \sqrt{1 - ( Z\alpha )^2 }$.

\ni
Note the large "spin crisis" $ \Delta q = 1/3 $ for $Z\alpha = 1$.

\subsection{ Results for the polarized densities in $(\bv, k^+ ) $}

The 2-component wave functions of a $ j_z = + 1/2 $ 
state write
\begin{equation} 
% \Phi_{\uparrow } 
\Phi(\bv, k^+ ) = 
\left( 
\begin{array}{c}
w \\ 
-iv \ e^{i\phi} \\ 
\end{array}
\right)
\,,\qquad
\Phi(\kv_T, k^+ ) = 
\left( 
\begin{array}{c}
\tw \\ 
- \tv \ e^{i\phi} \\ 
\end{array}%
\right)
\,.
\end{equation}
Other orientations of $ \Sv^A $ can be obtained by 
rotation in spinor space.

For the $(\bv, k^+)$ representation we can ignore the 
second term of (\ref{chi}). Then 
$v(\bv,k^+) $ and $w(\bv,k^+) $ are real and given by
\begin{equation}
%\begin{pmatrix} %????
\pmatrix{ 
v \cr w
}
%\end{pmatrix}
= \int_{-\infty}^{\infty } dz \ 
%\begin{pmatrix} %????
\pmatrix{ 
\xi b  \cr 
r + i \xi z 
}
%\end{pmatrix}
\ e^{-i k^+ z +i E z -i \chi(\bv,z)} \ f(r)/r 
\end{equation}
where $f(r) \propto r^{\gamma - 1} \ \exp(-m Z\alpha r) $ 
is the radial wave function. Then,
%~\cite{Landau, Bj&Dr, Akh&Ber, Mess}. Then
\begin{eqnarray}
q(b, k^+) &=& w^2 + v^2  \nonumber \\
C_{nn} (b, k^+) &=& 1 \nonumber  \\
C_{0n} (b, k^+)  = C_{n0} (b, k^+)  
&=& - 2 \ wu / (w^2 + v^2 ) \nonumber  \\
C_{ll} (b, k^+)  = C_{\pi \pi } (b, k^+) 
&=& ( w^2 - v^2 ) / (w^2 + v^2 ) \nonumber \\
C_{l \pi} (b, k^+)  = C_{ \pi l} (b, k^+) &=& 0
\,.
\label{Cij(b)}
\end{eqnarray}

\subsubsection{Sum rule for the atom magnetic moment}
Consider a classical object at rest, of mass $M$, 
charge $Q$, spin $\Jv$ and time-averaged magnetic moment 
$\vec\mu$. 
% and with spin $\Jv$ along $\xu$.
Its center of mass $\rv_G$ and the average 
center of charge $\langle \rv_C \rangle$ coincide, say at $\rv = 0$.
Upon a boost of velocity $\vv$, the center of energy $\rv_G$ 
and $\langle \rv_C \rangle$ are displaced laterally by 
\begin{equation}
\bv_G = \vv \times \Jv /M
\,,\quad 
\langle \bv_C \rangle = \vv \times \vec \mu /Q
\,.
\end{equation}
$\bv_G $ and $ \langle \bv_C \rangle $ coincide if the gyromagnetic ratio 
has the Dirac value $Q/M$. For the hydrogen atom, $ \bv_G $ is 
negligible and the magnetic moment is almost fully anomalous.
In the infinite momentum frame ($\vv \simeq \zu$), 
we have an electric dipole moment~\cite{Burkardt} 
\begin{equation}
-e \ \langle \bv \rangle  
= \mu_A \ \zu \times \bar \Sv^A 
\,,
\end{equation}
the transverse asymetry of the $\bv$ distribution coming from
the $ C_{0n} $ term of (\ref{densPol}). 
We recover the atom magnetic moment 
\begin{equation}
\mu_A = - e \ (1+2\gamma)/(6m_e)
\end{equation}
(the anomalous magnetic moment of the electron being omitted).

\subsection{ Results for the polarized densities in $(\kv_T, k^+ ) $}

For the $(\kv_T, k^+)$ representation we should take 
$z_0 = \pm \infty $ but then (\ref{chi}) diverges. 
In practice we assume some screening of the Coulomb potential and 
take $|z_0|$ large but finite, which gives 
\begin{equation}
\chi(\bv,z) = 
-Z\alpha \ \left[ \ \pm \ln(2 |z_0|/b) + \sinh^{-1} (z/b) \ \right]
\,,
\end{equation}
the upper sign corresponding to Compton scattering 
and the lower sign to annihilation.
Modulo an overall phase, 
\begin{equation}
%\begin{pmatrix} %????
\pmatrix{ 
\tw \cr \tv
}
%\end{pmatrix}
= 2\pi \int_0^\infty b \ db \ b^{\mp iZ\alpha} \  
%\begin{pmatrix} %???? 
\pmatrix{ 
J_0 (k_T b ) \ w (b,k^+) \cr
J_1 (k_T b ) \ v (b,k^+)
}
%\end{pmatrix}
\,,
\end{equation}
\begin{eqnarray}
q( k_T , k^+) &=& |\tw|^2 + |\tv|^2  
\nonumber \\
C_{nn} (k_T , k^+) &=& 1 \nonumber  \\
C_{0n} (k_T , k^+)  = C_{n0} (k_T , k^+)  
&=& 2 \Im (\tv^*\tw) / (|\tw|^2 + |\tv|^2 )  
\nonumber \\
C_{ll} (k_T , k^+)  = C_{\pi \pi } (k_T , k^+) 
&=& ( |\tw|^2 - |\tv|^2 ) / (|\tw|^2 + |\tv|^2 ) \nonumber  \\
C_{l \pi} (k_T , k^+)  = - C_{ \pi l} (k_T , k^+) 
&=& 2 \Re (\tv^*\tw) / (|\tw|^2 + |\tv|^2 ) 
\,. 
\label{Cij(k)}
\end{eqnarray}

The factor $ b^{- iZ\alpha} $ (Compton case)  
behaves like a converging cylindrical wave. Multiplying
$\Phi(\rv)$, it mimics an additional momentum 
of the electron toward the $z$ axis. 
In fact it takes into account the "focusing" of the final particle 
by the Coulomb field~\cite{Burkardt}. 
It also provides the relative phase between $\tw$ and $\tv$
which gives non-zero $C_{0n} (k_T , k^+) $ and $C_{n0} (k_T , k^+) $ 
(Sivers and Boer-Mulders-Tangerman effects).
% (\ref{sivers}) 
Similarly, $ b^{+ iZ\alpha} $ (annihilation case) takes into account
the defocusing of the incoming positron. 

\subsection{\protect\bigskip Positivity constraints}

The spin correlations between the electron and the atom can be encoded 
in a "grand density matrix" $R$~\cite{Artru-Richard}, 
which is the final density matrix of the crossed reaction 
$nucleus \to atom(\Sv^A) + e^+(-\Sv^e) $.

Besides the trivial conditions $| C_{ij} | \le 1$ 
the positivity of $R$ gives
\begin{equation}
( 1 \pm C_{nn} )^2  \ge 
(  C_{n0}  \pm C_{0n}  )^2 +
(  C_{ll} \pm  C_{\pi\pi} )^2 +
(  C_{\pi l}  \mp  C_{l\pi} )^2 
\,
\label{cond}
\end{equation}
These two inequalities as well as $| C_{ll} | \le 1$ 
are saturated by (\ref{Cij(b)}) and (\ref{Cij(k)}). 
In fact $R$ is found to be of rank one. 
It means that the spin information of the atom 
is fully transferred to the electron, once
the other degrees of freedom ($k^+$ and $\bv$ or $\kv_T$) 
have been fixed. If there is additional electrons or 
if we integrate over $k^+$, for instance, 
some information is lost and some positivity conditions
get non-saturated.
%(this is not the case for $ C_{nn} = 1$, which is a linear relation).
Conversely, the hypothesis that $R$
is of rank one leaves only two possibilities:
\begin{equation}
C_{nn} = \pm 1 
\,,\quad
C_{0n} = \pm C_{n0} 
\,,\quad
C_{\pi \pi } = \pm C_{ll} 
\,,\quad
C_{l \pi} = \mp  C_{ \pi l}
\,,
\end{equation}
with (\ref{cond}) saturated. 
The hydrogen ground state choses the upper sign.
After integration over $\bv$ or $\kv_T$, we are left with the 
{\it Soffer inequality}, 
\begin{equation}
2 \ | \delta q (k^+) | \le q (k^+) + \Delta q (k^+) 
\label{Soffer}
\end{equation}
which in our case is saturated, even after integration over $k^+$,
see (\ref{crise}).
 
\section{\protect\bigskip The electron-positron sea }
The charge rule (\ref{V-charge}) involves positive 
contributions of both positive and negative values of $k^+$.   
So it seems that there is less than one 
electron (with $k^+ >0$) in the atom.  
This paradox is solved by the introduction of the electron-positron sea. 

Let $ | n \rangle $ be an electron state in the Coulomb field.
Negative $n$'s are assigned to negative energy 
scattering states.
Positive $n$'s up to $ n_B$ label the bound states ($ -m <E_n < + m $) 
and the remaining ones from $ n_B+1$ to $ + \infty $ 
are assigned to unbound states of positive energy, $ E_n > + m $,
considered as discrete.  
Let $ | k,s \rangle $ be the plane wave of 
four-momentum $k$ and spin $s$, solution 
the free Dirac equation. 
The destruction and creation operators in these two bases
are related by
\begin{equation}
\alpha_{k,s}  = \sum_n \ \langle k,s | n \rangle \ a_n 
\,,\qquad
\alpha^\dagger_{k,s}  = \sum_n \ a^\dagger_n \ \langle n | k,s \rangle
\,. 
\label{passage}
\end{equation}
In the Dirac hole theory, the hydrogen-like atom is in the Fock state
\begin{equation}
| H_n \rangle = a^\dagger_n \ a^\dagger_{-1} \ a^\dagger_{-2} 
\cdots a^\dagger_{-\infty} 
\ | \hbox{QED-bare nucleus} \rangle 
\label{Fock}
\end{equation}
The number of electrons in the state $ | k,s \rangle $ is
\begin{equation}  
N^{elec} (\kv,s) = \langle \alpha^\dagger_{k,s} \ \alpha_{k,s} \rangle  
= | \langle k,s | n \rangle  |^2 
+ \sum_{n'<0}| \langle k,s | n' \rangle  |^2
\label{elec/atom}
\end{equation}
For a nucleus alone (but "QED-dressed")
the first factor $a^\dagger_n$ of (\ref{Fock}) is missing
and the first term of (\ref{elec/atom}) is absent. Therefore,
passing to the continuum limit
$ \langle k,s | n \rangle \ \longrightarrow \ \Phi ( \kv_T , k^+ ) $,
\begin{equation}  
N^{elec} _{atom} - N^{elec} _{nucleus} = \sum_{k, k^0>0} \ \sum_s 
\ | \langle k,s | n \rangle  |^2 
= \int_{k^+ >0} {dk^+ \over 2 \pi} \ q ( k^+ ) 
\,.
\label{k+pos}
\end{equation}

One term $n'$ of (\ref{elec/atom}) corresponds, e.g., 
to Compton scattering on an electron of the Dirac sea, 
producing a fast electron plus the positron  
$| n' \rangle $ = vacant. It is the {\it fracture function}
of reaction $ \gamma_i + A \to \gamma_f + A + e^+ + X $.

Similarly the number of positrons is 
(with $k^0 >0$): 
\begin{equation}  
N^{posit} (\kv,s) = 
\langle \alpha_{-k,-s} \ \alpha^\dagger_{-k,-s}  \rangle
= \sum_{0 < n' \ne n}| \langle -k,-s | n' \rangle  |^2
\,.
\label{posit/atom}
\end{equation} 
For a nucleus alone the condition $n' \ne n$ is relaxed. 
By difference, 
\begin{equation}
N^{posit} _{nucleus} - N^{posit} _{atom} 
= \int_{k^+ < 0} {dk^+ \over 2 \pi} \ q ( k^+ ) 
\label{posit/at-nuc}
\end{equation}
One term of Eq.(\ref{posit/atom}) corresponds to the extraction of 
an electron of large negative energy, giving a 
fast positron and an electron in $| n' \rangle $. 
Eqs.(\ref{k+pos}) + (\ref{posit/at-nuc}) and (\ref{V-charge})
tell that the atom and nuclear charges differ by $e$. 
But the charge of the electron-positron virtual cloud 
surrounding the nucleus is not zero (charge renormalisation). 

\section{CONCLUSION}

We have seen that the leading twist structure functions of the 
hydrogen-like atom at large $Z$ has many properties that 
are supposed or verified for the hadronic ones, in particular: 
sum rules, longitudinal "spin crisis", Sivers effect, 
transverse electric dipole moment in the $P_\infty$ frame, etc. 
It remains to evaluate these effects quantitatively. 
With this "theoretical laboratory" 
one may also investigate spin effects in fracture functions,
non-leading twist structure functions,
Isgur-Wise form factors, etc.  
The electron-positron sea may deserve further studies:
to what extent is it polarized or asymmetrical in charge ?
Is the charge renormalisation of the nucleus found in Sec.6 
the same as in standard QED ? Is it finite and calculable 
for an extended nucleus ?

Finally it may be interesting to do or re-analyse experiments 
on deep inelastic Compton scattering ("Compton profile" measurements).

 %\begin{center}
%\setcounter{page}{12}{\Large \addcontentsline{toc}{chapter}{R��ences}}
%\end{center}

\end{document}